\documentclass[review]{elsarticle}
\usepackage{booktabs} 
\usepackage{graphicx}
\usepackage{amsfonts}
\usepackage{dsfont}
\usepackage{graphicx, epsfig,amsmath,amssymb,latexsym,graphics,float}
\usepackage{setspace,subfig}
\usepackage{citesort}
\usepackage{float}
\usepackage[ruled,vlined,linesnumbered]{algorithm2e}
\usepackage{booktabs}
\usepackage{lipsum}
\usepackage{multicol}
\usepackage{url}
\usepackage{algorithmic}
\def\x{{\mathbf x}}

\begin{document}
\begin{frontmatter}
\title{A Survey on Trust Modeling from a Bayesian Perspective
}
\author{Bin Liu$^{1,2}$}
\address{$^{1}$ School of Computer Science, Nanjing University of Posts and Telecommunications\\
$^{2}$ Jiangsu Key Lab of Big Data Security $\&$ Intelligent
Processing\\ Email: bins@ieee.org}
\begin{abstract}
In this paper, we are concerned with trust modeling for agents in
networked computing systems. As trust is a subjective notion that is
invisible, implicit and uncertain in nature, many attempts have been made to model trust with aid of Bayesian probability theory, while the field lacks a global comprehensive analysis for variants of Bayesian trust
models. We present a study to fill in this gap by giving a
comprehensive review of the literature. A generic Bayesian trust (GBT) modeling perspective is highlighted here. It is shown that all models under survey can cast into a GBT based computing paradigm as special cases. We discuss both capabilities and limitations of the GBT perspective and point out open questions to answer, with a hope to advance GBT to become a pragmatic infrastructure for analyzing intrinsic relationships among variants of trust models and developing novel tools for trust evaluation.
\end{abstract}

\end{frontmatter}
\section{Introduction}
In the past decade, we have witnessed the advent of a variety of networked computing systems.
Examples include wireless sensor networks (WSNs), internet of things (IoT), electronic commerce (EC),
P2P networks, cloud computing, mobile ad hoc networks (MANETs), cognitive radio networks (CRNs), multi-agent systems, semantic web, social networks,
web-based recommender systems. A basic feature of such systems is the exploitation of the collective power of the network nodes to
accomplish novel tasks.

In tandem with advances in network technologies, this has spurred a
lot of renewed interest in trust modeling and evaluation. Generally
speaking, trust is a concept with regard to the expectation, belief,
and confidence on the integrity, reliability, security,
dependability, ability, and other characters of an entity. It is a
key factor that influences collaboration, competition, interaction
and information sharing among network peers, and an enabling
technology for decision making that facilitates the achievement of a system
or application goal. We will use the words agents, entities, nodes,
principals or peers interchangeably through the rest of this paper,
as we will do with interactions, relationships, and links.

Trust has been studied for a long time \cite{mcknight1996meanings}.
The starting point of these studies may originate from the social
sciences, for which trust between humans and its effects in economic
transactions are the focus
\cite{ba2002evidence,gambetta1988trust,mcknight1996trust}. The
notion of trust is intuitively easy to comprehend, while it has not
yet been formally defined.
Various definitions of trust have been proposed in different computing fields.
Most of them depend on the context in which trust is investigated or the viewpoint adopted.
Here are some examples as follows.
\begin{quote}
  ``Trust is the subjective probability by which an individual, A, expects that another individual, B, performs a given action on which
its welfare depends. '' \cite{gambetta2000can}
\end{quote}
\begin{quote}
  ``Trust is the extent to which one party is willing to
depend on something or somebody in a given situation with a feeling of relative
security, even though negative consequences are possible.'' \cite{mcknight1996meanings}
\end{quote}
\begin{quote}
  Trust is ``a belief that is influenced by the individual's
  opinion about certain critical system features.'' \cite{kini1998trust}
\end{quote}
\begin{quote}
 ``Trust (or, symmetrically, distrust) is a particular level of the
subjective probability with which an agent assesses that another agent or group of
agents will perform a particular action, both before he can monitor such action (or
independently of his capacity ever to be able to monitor it) and in a context in
which it affects his own action.'' \cite{gambetta2000can}
\end{quote}
\begin{quote}
 ``Trust is a psychological state comprising the intention to accept vulnerability
 based upon positive expectations of the intentions or behavior of another.'' \cite{Rousseau1998}
 \end{quote}
\begin{quote}
  ``Trust is the willingness of the trustor (evaluator) to take risk based on a subjective
belief that a trustee (evaluatee) will exhibit reliable behavior to maximize the trustor's
interest under uncertainty (e.g., ambiguity due to conflicting evidence and/or ignorance
caused by complete lack of evidence) of a given situation based on the cognitive assessment
of past experience with the trustee.'' \cite{cho2015survey}
\end{quote}
\begin{quote}
``Trust is a phenomenon that humans use every day to promote interaction and accept risk in situations where only partial information is
available, allowing one person to assume that
another will behave as expected.'' \cite{cahill2003using}
\end{quote}

In addition to existing scattered definitions of trust, models for
expressing trust involved in networked systems also lack coherence
and consistency. Examples of the main formal techniques for modeling
trust include fuzzy logic
\cite{selvaraj2017evidence,rafique2016black,nagy2008multi,lesani2006fuzzy,chen2009comparison,liao2009fuzzy,luo2008fuzzy,manchala1998trust,nefti2005fuzzy},
subjective logic
\cite{josang2016bayesian,jiang2015efficient,filali2015global,alhadad2014trust,ahmadi2015probabilistic,cerutti2015subjective,liu2014assessment,josang1999algebra,josang2001logic,lioma2010subjective,oren2007subjective},
Dempster-Shafer theory
\cite{deepa2014trust,wang2007inverse,wang2007trust,zhang2017novel,nguyen2016integrating,esposito2018information},
ratings
\cite{resnick2000reputation,abdul2000supporting,jonker1999formal,jonker2004human,buchegger2004robust},
weighting
\cite{pirzada2004establishing,sabater2001regret,wang2005two,azzedin2002evolving,hung2007trust},
neural network \cite{song2004neural,baohua2005identifying}, Bayesian
networks
\cite{songsiri2006mtrust,wang2006bayesian,momani2008bnwsn,nguyen2007bayesian},
game theory \cite{michiardi2002core,xiong2003reputation}, swarm
intelligence
\cite{jiang2004ant,wang2006ant,marmol2011providing,marmol2009tacs},
credential and policy
\cite{santos2012policy,neuman1994kerberos,winslett2002negotiating,li2003distributed,nejdl2004peertrust,bonatti2005driving,winsborough2000automated,becker2004cassandra,olmedilla2007security},
and others
\cite{lee2001trust,theodorakopoulos2006trust,bao2012hierarchical,boukerche2008trust,kamvar2003eigentrust,xiong2004peertrust,sabater2001regret,regan2005model,zacharia2000trust,su2010pbtrust}.

Despite the discrepancy in definitions of and modeling tools for
trust, it is well recognized that trust is important for promoting
quick responses to a crisis, reducing transaction costs, avoiding
harmful conflict, and enabling cooperative behavior. Trust can be
regarded as a measure of a peer's faithfulness. It can also be
treated as a prediction of the future behaviors of the peer who
provide service. The trust value is thus a probability that
a peer behaves satisfactorily in future interactions.
When multiple aspects of the peer's interaction with other peers are
considered, a trust vector, instead of a single value, can be used.
Each element of this vector is used for evaluating one aspect. Most
of the existing work on trust is highly specific to application
considered. The theoretical underpinning of trust modeling and
evaluation is important for creating a more cumulative body of
knowledge on trust, while it remains a hard nut to crack.


In this paper, we focus on a Bayesian perspective for trust modeling for
agents in networked computing systems. In spite of a few attempts to model trust using Bayesian techniques
\cite{zouridaki2007trust,liu2015toward,liu2015probabilistic,Denko2011trust,liu2016state,chen2014dynamic,wang2003bayesian,dubey2014bayesian,venanzi2015bayesian,xu2015optimo},
the field lacks a global overview for variants of Bayesian trust
models and a discussion on their connections between each other. We present a
study to fill in this gap.
This paper does not pretend to be an exhaustive bibliography survey,
but rather will review a gallery of selective research. A comprehensive bibliography can be found
in other survey papers such as
\cite{guo2017survey,yan2014survey,josang2007survey,zhang2011survey,cho2015survey,viljanen2005towards,artz2007survey,yu2010survey,momani2010survey,momani2010trust,grandison2000survey,sherchan2013survey,pinyol2013computational,lopez2010trust}.

The major contributions of of this paper are summarized as follows.
\begin{enumerate}
\item We provide a high-level generic Bayesian perspective, termed generic Bayesian trust (GBT) here, for trust modeling and evaluation;
\item We give a literature review of classic variants of Bayesian trust models and some alternatives which can cast
into the GBT paradigm;
\item We identify the strengths and weaknesses of the GBT paradigm;
\item We make an attempt to provide an improved understanding of Bayesian philosophy as well as its capabilities and limitations in modeling trust.
\end{enumerate}
\subsection{Organization of this paper}
The remainder of this paper is organized as follows. Section 2
presents the GBT perspective and basic mathematical tools required
to implement it. Section 3 provides a literature review of trust
models that leverage the Beta distribution and Dirichlet
distribution as the cornerstone. The relationship between these
models and the GBT perspective is also discussed. Section 4 reviews
state-space based trust models and discusses their connections to
the GBT perspective. Section 5 presents a literature review on the
subjective logic based trust models, and discusses their
relationships to the GBT perspective. Section 6 provides an overall comparison of
all models under survey. Section 7 discusses the
capabilities and limitations of the GBT perspective in trust
modeling and points out a number of open research questions.
Finally, Section 8 concludes the paper.
\section{The GBT Perspective for Trust Modeling and Evaluation}\label{sec:gbt}
The GBT perspective is based on a collection of ideas.
Assume that a trustor principal is interested in but uncertain about
the trust (or distrust) of a trustee principal in a specific
context. The trustor can quantify its uncertainty as a probability
for the quantity it is interested in, and as a conditional
probability for observations it might make, given the quantity it is
interested in. When data arrives, Bayes theorem tells the trustor
how to move from its prior probability to a new conditional
probability for the quantity of interest, also known as the
posterior probability in the jargon of Bayesian statistics. In this
view, trust (or distrust) is quantitatively measured with a
probability value ranging from 0 to 1.

The major insight leveraged in the GBT perspective is that, in spite
of the complexity regarding the concept of trust, it has two natural
and intrinsic attributes, namely uncertainty and subjectivity. The
GBT perspective takes into account of such attributes as fully as
possible to grasp the essence of trust, rendering it compatible with
and applicable across different application domains.

To begin with, let us consider an open-ended network, which consists of network agents and links among the agents.
At any point in time, there may be new agents joining the network and/or some others leaving.
Each agent is associated with a unique identity (ID) and may interact with some others at some point in time.
When a pair of agents interact with each other, we say that there is a link between them.
Such interactions may take place between different pairs of network agents over time.
During an interaction between a pair of agents, one agent plays the role of the trustor, and the other is called trustee.
We can regard the trustee as the agent obliged to provide a service, and trustor as that receiving this service.
The notion of trust is interpreted as the trustor's expectation of a certain future behavior of the trustee based on
first-hand experience in interacting with it in the past and other possiblely relevant information provided by third-party agents.

We use the random variable $\Theta$ to denote the probabilistic trust of the trustee from the point of view of the trustor and use $Y$ to represent data the trustor observes. Realizations of $\Theta$ and $Y$ are respectively denoted by $\theta$ and $y$.
We consider data items sequentially and represent the data item arriving at the $k$th time step by $y_k$.
As usual, the symbol $t$ is used as the continuously valued time variable.
The value of $t$ at the $k$th time step step is $t_k$.

Taking a Bayesian perspective, we translate the trust inference problem to recursively calculate the degree of
belief in $\theta_k$ taking different values based on data observed up to time step $k$, namely $y_{1:k}\triangleq\{y_1,\ldots,y_k\}$.
Specifically, we need to derive the probability density function (pdf) of $\theta_k$ conditional on observations $y_{1:k}$.
This pdf is termed the posterior and represented as $p(\theta_k|y_{1:k})$ or $p_{k|k}$ in short.
Denote the initial pdf, also known as the prior, of the trust variable by $p(\theta_0|y_0)\equiv p(\theta_0)$, where $y_0$ denotes
the set of no measurements.
Then, the posterior can be obtained by recursively running two stages, namely, prediction and update.

Suppose that the posterior at time step $k-1$, namely $p(\theta_{k-1}|y_{1:k-1})$, is available.
The prediction stage runs based on the Chapman-Kolmogorov equation, yielding the prior pdf of the $\theta$ associated with time step $k$, as follows
\begin{equation}\label{eqn:pred}
p(\theta_k|y_{1:k-1})=\int p(\theta_k|\theta_{k-1})p(\theta_{k-1}|y_{1:k-1})d\theta_{k-1}.
\end{equation}
In the above equation, $p(\theta_k|\theta_{k-1})$ is determined by a transition function
\begin{equation}\label{eqn:trans}
\theta_k=\mbox{f}_k(\theta_{k-1},\mbox{v}_{k-1})
\end{equation}
where the function $\mbox{f}_k$ characterizes the time-evolution law of $\theta$ from time $t_{k-1}$ to $t_k$,
$\mbox{v}$ is an independent and identically distributed (i.i.d.) noise sequence used for modeling the uncertainty underlying the evolution law.

Upon the arrival of the new observation $y_k$, the update stage is invoked to update the prior pdf via Bayes' rule as follows
\begin{equation}\label{eqn:update}
p_{k|k}=\frac{p(y_k|\theta_k)p(\theta_k|y_{1:k-1})}{p(y_k|y_{1:k-1})}
\end{equation}
where the denominator is termed normalizing constant defined as
\begin{equation}\label{eqn:nc}
p(y_k|y_{1:k-1})=\int{p(y_k|\theta_k)p(\theta_k|y_{1:k-1})}d\theta_k,
\end{equation}
The likelihood function $p(y_k|\theta_k)$ here quantifies the trustor's uncertainty on the observation $y_k$, given the trust value $\theta_k$.
It is worth to note that the task here is to compute the expected value of the unknown $\theta_k$, for which we need to derive the posterior $p_{k|k}$,
but it is not necessary to compute the likelihood for any particular value of $\theta_k$. Based on Bayesian principle, $p_{k|k}$ can be computed from $p_{k-1|k-1}$ recursively as follows
\begin{equation}\label{eqn:filter}
p_{k|k}=\frac{p(y_k|\theta_k)\int p(\theta_k|\theta_{k-1})p_{k-1|k-1}d\theta_{k-1}}{p(y_k|y_{0:k-1})}.
\end{equation}

Except for a few special cases, there is no analytic closed-form solution to (\ref{eqn:filter}). Here we introduce an approximate method, namely particle filter (PF) \cite{arulampalam2002tutorial}, to solve (\ref{eqn:filter}). The PF plays the role of the major computational engine for implementing the GBT paradigm.
The PF algorithm has been widely used for tracking dynamic target distributions
\cite{liu2011instantaneous,arulampalam2002tutorial,liu2010multi,liu2013sequential,liu2008particle,liu2017robust,yi2016robust}.
Suppose that we have a set of weighted samples $\{\theta_{0:k-1}^i,\omega_{k-1}^i\}_{i=1}^N$ at time step $k-1$, which can be used to construct a discrete
approximation of $p_{k-1|k-1}$. The samples are drawn from a proposal density, namely $\theta_{0:k-1}^i\sim q(\theta_{0:k-1}|y_{1:k-1})$,
and the sample weights satisfies
$\omega_{k-1}^i\propto p(\theta_{0:k-1}|y_{1:k-1})/q(\theta_{0:k-1}|y_{1:k-1})$, $\sum_{i=1}^N\omega_{k-1}^i=1$.
The proposal density $q(\theta_{0:k-1}|y_{1:k-1})$ is chosen to factorize such that
\begin{equation}
q(\theta_{0:k}|y_{1:k})=q(\theta_k|\theta_{0:k-1},y_k)q(\theta_{0:k-1}|y_{1:k-1}).
\end{equation}
Each sample corresponds to a trajectory of the trust values up to time step $k-1$.
At time $k$, each trajectory, say the $i$th one, extends by sampling $\theta_k^i$ from a proposal distribution $q(\theta_k|\theta_{k-1}^i,y_k)$ and then being weighted by
\begin{equation}\label{eqn:pf_weight}
\omega_k^i\propto\omega_{k-1}^ip(\theta_k^i|\theta_{k-1}^i)p(y_k|\theta_k^i)/q(\theta_k^i|\theta_{k-1}^i,y_k).
\end{equation}
The posterior density $p_{k|k}$ can then be approximated by the updated sample set as follows
\begin{equation}\label{eqn:pf_posterior_approx}
p_{k|k}\thickapprox\sum_{i=1}^N\omega_k^i\delta_{\theta_k^i},
\end{equation}
where $\delta_{\theta}$ denotes the Dirac-delta function located at $\theta$.

Starting from $\{\theta_{k-1}^i,\omega_{k-1}^i\}_{i=1}^N$, we summarize an iteration of the algorithm as follows:
\begin{itemize}
\item Sampling step. Sample $\hat{\theta}_k^i\sim q(\theta_k|\theta_{k-1}^i,y_k)$, $\forall i$;
\item Weighting step. Set $\omega_k^i$ using (\ref{eqn:pf_weight}), $\forall i$; Normalize them to guarantee that $\sum_{i=1}^N\omega_k^i=1$;
\item Resampling step. Sample $\theta_k^i\sim\sum_{j=1}^N\omega_k^j\delta_{\hat{\theta}_k^j}$, set $\omega_k^i=1/N$, $\forall i$.
\end{itemize}
\begin{algorithm}[tb]
\caption{A pseudo-code to implement trust inference based on the GBT perspective}
\label{alg:GBT}
\begin{algorithmic}[1] 
\STATE Initialize the prior density function.
\FOR{$k$=1,2,\ldots}
\STATE Compute the predictive distribution of the trust variable using Eqn.(\ref{eqn:pred});
\STATE Calculate the posterior given by Eqn.(\ref{eqn:update}) if an analytical solution to it exists; otherwise, run the PF algorithm to get a particle approximation of the posterior.
\STATE Set the prior density of the next iteration to be the (approximate) posterior at the above step.
\ENDFOR
\end{algorithmic}
\end{algorithm}
As presented above, using the GBT perspective we translate trust evaluation to be a Bayesian sequential inference problem. See Algorithm \ref{alg:GBT} for a pseudo-code to implement trust inference based on the GBT perspective. We will show in Sections 3-6 that several most classic trust models cast into the GBT perspective if we appropriately define the prior and the likelihood function.
\section{Beta distribution and Dirichlet distribution based trust models}\label{sec:review_bayes_models}
This section presents a review of trust models that leverage Beta
distribution and Dirichlet distribution models as major
cornerstones. The relationships between these models and the GBT
perspective presented in Section \ref{sec:gbt} is discussed in Section
\ref{sec:relation_bdtm}.
\subsection{Beta distribution based trust model (BDTM)}\label{sec:bdtm}
As its name indicates, the basic idea of BDTM is using a parametric Beta model to represent the distribution of the
probabilistic trust $\Theta$ \cite{josang2002beta,teacy2005coping,teacy2006travos,huynh2006integrated,patel2005probabilistic}.
The basic assumption leveraged by BDTM is that the outcome of each interaction is binary, e.g., $\{$success (good), failure (bad)$\}$.
The trust variable $\Theta$ is interpreted as the probability that the outcome of an
forthcoming interaction between the trustee and the trustor will be success (good).
Therefore, a sequence of $n$ interactions $\textbf{X}=X_1,\ldots,X_n$ is represented
as a sequence of binomial (Bernoulli) trials and then modeled by a binomial distribution
\begin{equation}\label{eqn:binomial}
p(\textbf{X}\,\mbox{consists of}\,m\, \mbox{successes})=\theta^m(1-\theta)^{n-m}.
\end{equation}
We can see that if the prior of $\Theta$ is represented by a Beta distribution
$B(\alpha,\beta)\propto\theta^{\alpha-1}(1-\theta)^{\beta-1}$, then
the posterior $p(\Theta|\textbf{X})$ will be $p(\Theta|\textbf{X})$
is $B(\alpha+m,\beta+n-m)$, where $m$ denotes the number of
successes in $\textbf{X}$, $\alpha$ and $\beta$ are parameters of the prior. See Algorithm \ref{alg:BDTM} for a pseudo-code to implement trust inference based on the BDTM.
\begin{algorithm}[tb]
\caption{A pseudo-code to implement trust inference based on the BDTM}
\label{alg:BDTM}
\begin{algorithmic}[1] 
\STATE Initialize the prior density function with a Beta distribution $B(\alpha,\beta)$.
\FOR{$k$=1,2,\ldots}
\STATE Analyze the new observation data $y_k$, which consists of a sequence of $n$ interactions $\textbf{X}=X_1,\ldots,X_n$. Count the number $m$ of successes in $\textbf{X}$;
\STATE Update the posterior $B(\alpha,\beta)$ by setting $\alpha\leftarrow\alpha+m, \beta\leftarrow\beta+n-m$.
\ENDFOR
\end{algorithmic}
\end{algorithm}

In real applications of BDTM, a disturbing issue may exist due to
the so-called whitewashing problem, namely, the trustor has little
or no previous experience in interacting with the trustee. A
pragmatic solution to such problem is to allow the trustor to
`consult' about the trustee's behavior from third-party peers
\cite{Teacy2008Hierarchical,liu2015probabilistic}, and then
integrate its direct experience with information reported by
third-party peers when assessing trust in the trustee.

A robust multiagent system is demanded to be able to deal with selfish, antisocial, or unreliable agents.
A model presented in \cite{Teacy2008Hierarchical} characterizes the relationship between individual peer behaviors and group behaviors of peers
through a two-layered hierarchy. Liu and Yang take into account of the possibility of that the third-party peers may be unreliable and thus
reports provided by them may be
inaccurate \cite{liu2015probabilistic}. A notion, termed Advisor-to-Trustor Relevance (ATTR) metric,
is proposed in \cite{liu2015probabilistic} to quantitatively measure the systematic correlation of the subjectivity of the trustor and that
of the third-party peer (i.e.,
the so-called advisor in \cite{liu2015probabilistic}), based on the assumption that there is a common set of trustees, with which both the trustor and the advisor have interacted in the past. A report given by an advisor is weighted by its corresponding ATTR measure and then is used to update the posterior of $\Theta$ on
the basis of BDTM \cite{liu2015probabilistic}.

To get rid of performance deterioration due to the presence of
unfair rating provided by the advisors, specific data processing
mechanisms are designed to filter unfair ratings that may result in
misleading trust evaluations when using BDTM
\cite{whitby2004filtering}. One major mechanism is to consider
recent ratings more by `forgetting' old ratings as in the Bayesian reputation system (BRS) \cite{josang2016bayesian}. The
basic assumption leveraged by BRS is that opinions provided by the
majority of reputation sources are accurate, so any opinions
deviating significantly from the average will be ignored. In
contrast with BRS, the TRAVOS (Trust and Reputation model for Agent
based Virtual OrganisationS) \cite{teacy2005coping,teacy2006travos}
copes with unfair ratings by learning to distinguish reliable from
unreliable third-party peers through repeated interactions with
individual peers. However, TRAVOS relies on the assumption that the trustor and the third-party peers
have extensive historical interactions that enable each third-party
peer's expected honesty to be assessed. In addition, there is no
time discounting for reports provided by third-party peers in TRAVOS.
In contrast with TRAVOS, the Personalized Trust Model (PTM)
\cite{zhang2008evaluating} includes a forgetting factor to discount
less recent ratings given by third-party peers.
\subsection{Dirichlet distribution based trust model (DDTM)}\label{sec:ddtm}
The DDTM generalizes BDTMs with only `binary' outcomes to multiple typed outcomes \cite{mui2002computational,teacy2005coping,despotovic2006p2p,nielsen2007bayesian,reece2007rumours}.
Such outcomes can be interpreted as different degrees of success on the `success'–-`failure' scale.
In this case, $\Theta$ is in the form of a vector consisting of, say $b$ elements, viz. $\Theta\equiv[\Theta_1,\ldots,\Theta_b]$,
with $\Theta_i$ representing the probability that the $i$th type outcome in the $b$-way choice will happen in the next interaction.
Correspondingly, the multinomial distribution is used to model a $n$-sequence of trials $\textbf{X}$ with $b$ distinct outcomes.
As a conjugate prior to multinomial trials, the Dirichlet distribution,
\begin{equation}\label{eqn:dirichlet}
D(\alpha_1,\ldots,\alpha_b)\propto\theta_1^{\alpha_1-1}\dots\theta_b^{\alpha_b-1},
\end{equation}
is leveraged in DDTM, and then the posterior
$p(\Theta|\textbf{X})$ is also a Dirichlet distribution as follows
\begin{equation}
D(\alpha_1+\sharp_1(\textbf{X}),\ldots,\alpha_b+\sharp_b(\textbf{X})),
\end{equation}
where $\sharp_i(\textbf{X})$ counts the occurrences of the $i$th type outcome in the sequence $\textbf{X}$. See Algorithm \ref{alg:DDTM} for a pseudo-code to implement trust inference based on the DDTM.
\begin{algorithm}[tb]
\caption{A pseudo-code to implement trust inference based on the DDTM}
\label{alg:DDTM}
\begin{algorithmic}[1] 
\STATE Initialize the prior density function with a Dirichlet distribution $D(\alpha_1,\ldots,\alpha_b)$.
\FOR{$k$=1,2,\ldots}
\STATE Given a new observation $y_k$, which is a $n$-sequence of trials $\textbf{X}$ with $b$ distinct outcomes, count the occurrences of each outcome type in the sequence $\textbf{X}$;
\STATE Update the posterior $D(\alpha_1,\ldots,\alpha_b)$ by setting $\alpha_i\leftarrow\alpha_i+\sharp_i(\textbf{X})$ for $i=1,\ldots,b$.
\ENDFOR
\end{algorithmic}
\end{algorithm}
The DDTM has been widely used in the context of global ubiquitous computing \cite{nielsen2007bayesian}, especially E-Marketplaces \cite{regan2006bayesian}.
\subsection{On relationships between BDTM, DDTM and the GBT perspective}\label{sec:relation_bdtm}
In this section, we clarify relationships between BDTM, DDTM and the
GBT perspective.

If we define the state transition function (\ref{eqn:trans}) in GBT
as follows
\begin{equation}\label{eqn:trans2}
\theta_k=\theta_{k-1},
\end{equation}
it leads to $p(\theta_k|\theta_{k-1})=\delta_{\theta_{k-1}}$, and then $p(\theta_k|y_{1:k-1})=p(\theta_{k-1}|y_{1:k-1})$. That is to say that the prior pdf at time step $k$ is equivalent to the posterior at time step $k-1$. Now further restrict the initial pdf of $\theta$, $p(\theta_0|y_0)$, to be a Beta distribution, and then define the likelihood function $p(y_k|\theta_k)$ in (\ref{eqn:filter}) to be binomial as follows
\begin{equation}\label{eqn:binomial2}
p(y_k|\theta_k) = \theta_k^m(1-\theta_k)^{n-m},
\end{equation}
where $n$ denotes the number of binary data items included in $y_k$
and $m$ the number of successes in $y_k$. Here $y_k$ is assumed to
include multiple data items. Note that the GBT perspective allows any
number of interactions to be performed between a pair of peers
during a time interval. The resulting posterior (\ref{eqn:filter})
is then restricted to be a Beta distribution, due to the fact that the
class of Beta prior distributions is conjugate to the class of
binomial likelihood functions. Now the GBT framework reduces to the
BDTM.

To summarize, the BDTM can be regarded as a special implementation
case of the GBT framework, with specific definitions of the initial
pdf $p(\theta_0)$ (i.e., in the form of Beta distribution), state
transition function (as shown in (\ref{eqn:trans2})) and the
likelihood (as shown in (\ref{eqn:binomial2})).

Analogously, if we model the initial pdf $p(\theta_0)$ to be a
Dirichlet distribution, specify the transition function in the same
way as in (\ref{eqn:trans2}) and represent the likelihood function
to be multinomial, then the GBT framework reduces to the DDTM
presented in Subsection \ref{sec:ddtm}. Therefore, the DDTM can also
be regarded as an ad hoc implementation of the GBT framework.

In case of the trustee's past behavior data given by a third-party
peer being available, the GBT framework allows the behavior data
itself as well as the information on the third-party peer to be
treated as a new observation that is then used to update the posterior
pdf of $\Theta$. To handle such an observation using GBT, only an
appropriate likelihood function is required to be set. Suppose that
the Beta distribution model is under use and the trustee's past
behavior data $\mbox{X}$ reported by this third-party peer is an
$n$-sequence of $m$ successes, then the likelihood function can be
defined to be $\theta^{\theta_{tpp}m},\theta^{\theta_{tpp}(n-m)}$,
where $\theta_{tpp}$ denotes the trust value of this third-party
peer from the point of view of the trustor.  $\theta_{tpp}$ may be
calculated based on past interactions between this third-party peer
and the trustor. Then the resulting posterior of $\Theta$, after
seeing this new observation, becomes
$B(\alpha+\theta_{tpp}m,\beta+\theta_{tpp}(n-m))$, assuming that the
prior of $\Theta$ is $B(\alpha,\beta)$. If a series of new
observations from multiple third-party peers are available, then the
observations can be processed one by one, in the same way as
illustrated above, and, correspondingly, the posterior pdf will be
updated in a sequential manner.

Note that, for all cases discussed above, there is no need to apply
the PF sampling approach to calculate the posterior pdf as shown in
(\ref{eqn:filter}), because analytical solutions to
(\ref{eqn:filter}) are available due to the adopted conjugate prior
to the likelihood function.
\section{State-Space based trust model (SSTM)}\label{sec:sstm}
In contrast with the BDTM and DDTM that completely neglect the time-evolution feature of trust,
the SSTM characterizes the time-evolution law of trust directly as follows \cite{liu2016state,liu2015toward,wang2017online}
\begin{equation}\label{eqn:ss_transition}
\theta_k\sim\mathcal{T_N}(\alpha\theta_{k-1},Q),
\end{equation}
where $\mathcal{T_N}(m,Q)$ denotes a truncated normal pdf with mean $m$, variance $Q$ and support area $[0,1]$, $0\leq\alpha\leq1$ a forgetting factor specified by the model designer. An empirically setting of $\alpha$ is 0.85 given in \cite{liu2015toward}.
In SSTM, the data $y$ is a continuously valued vector represented as $y_k\triangleq[y_{k,0},y_{k,1},\ldots,y_{k,n_k}]$,
where $n_k$ denotes the number of neighbor (or similar) peers of the trustee at time step $k$. The first element $y_{k,0}$
denotes the interaction outcome observed by the trustor when interacting with the trustee.
The element $y_{k,i}, 0<i\leqslant n_k,$ denotes the outcome observed by the trustor when interacting with the $i$th neighbor peer of the trustee.
Then the relationship between $\theta_k$ and $y_k$ is formulated via a likelihood function
\begin{equation}\label{eqn:ss_likelihood}
p(y_k|\theta_k)=\exp\left(\frac{-\mid\theta_k-V\mid}{\beta}\right)
\end{equation}
where $0<\beta<1$ controls the degree of the sensitivity of the likelihood value with respect to $V$. The term $V$ denotes the averaged voting value over all neighbor peers, namely,
\begin{equation}\label{eqn:ss_voting}
V\triangleq \frac{\sum_{i=1}^{n_k} U(i,y_k)}{n_k},
\end{equation}
in which
\begin{equation}\label{eqn:ss_u}
U(i,y_k)=\left\{\begin{array}{ll}
1,\quad\mbox{if}\quad|y_{k,i}-y_{k,0}|< r \\
0,\quad\mbox{otherwise} \end{array} \right.
\end{equation}
where $r$ denotes the maximum allowable difference between $y_{k,0}$ and $y_{k,i}, i>0$ under the assumption that the trustee behaves in a trustworthy way.

If $U(i,y_k)=1$, it can be interpreted as the $i$th neighbor peer casting a vote of that the trustee is trusted. In the context of WSNs, the definition of $U(i,y_k)$ reflects a smooth variation in sensor readings of close-by reliable sensors \cite{liu2016state,liu2015toward,wang2017online}. A basic assumption leveraged here is that every neighbor peer of the trustee is totally trusted by the trustor.
This assumption is relaxed in \cite{liu2016state}, in which an iterative PF method is presented that takes account of the trustworthiness of every neighbor peer in the calculation of $V$. Specifically, in that case, $V$ is defined to be
\begin{equation}\label{eqn:weighted_V}
V\triangleq \frac{\sum_{i=1}^{n_k}\theta_{k,i}U(i,y_k)}{\sum_{i=1}^{n_k}\theta_{k,i}}.
\end{equation}
In a slight abuse of notation, here we use $\theta_{k,i}$ to denote the trust in the $i$th neighbor peer of the trustee, and use $\theta_{k,0}$ to denote the trust in the trustee; both are in the point of view of the trustor. It is worthy to note that every $\theta_{k,i}$, $i>0$ may be itself unknown. So an algorithm is required to estimate $\theta_k\triangleq[\theta_{k,0},\,\theta_{k,1},\,\cdots,\,\theta_{k,n_k}]$.
An iterative PF (IPF) approach is proposed in \cite{liu2016state} to estimate $\theta_{k,j}, j=0,1,\ldots,n_k$ one by one. The idea is to regard the $n_k+1$ network peers as members of a virtual committee. When estimating trust of a member in this committee, then this member is regarded as the trustee and the other members are regarded as neighbor peers of the trustee. A pseudo-code to implement IPF is presented in Algorithm \ref{alg:IPF}.
\begin{algorithm}[tb]
\caption{A pseudo-code to implement the IPF}
\label{alg:IPF}
\begin{algorithmic}[1] 
\STATE Initialize $\theta_{k,j}=1$, $j=0,1,\ldots,n_k$.
\FOR{$j=0, 1, \ldots, n_k$}
\STATE Treat the $j$th peer as the trustee and the others as the trustee's neighbor peers;
\STATE Given $\theta_{k,i}, i=0,\ldots,j-1,j+1,\ldots,n_k$, approximate the posterior pdf of $\theta_{k,j}$ by a weighted sample set obtained using one iteration of PF.
\STATE Update $\theta_{k,j}$ based on the weighted sample set.
\ENDFOR
\STATE If the stop criterion is satisfied, output $\theta_{k,j}, j=0,1,\ldots,n_k$; otherwise, return to step 2.
\end{algorithmic}
\end{algorithm}
\subsection{On relationship between SSTM and
the GBT perspective}\label{sec:relation_sstm}
The relationship between SSTM and the GBT perspective is easy to check. If we set the state
transition function (\ref{eqn:trans}) to be
(\ref{eqn:ss_transition}), and define the likelihood function
$p(y_k|\theta_k)$ as (\ref{eqn:ss_likelihood}), then the GBT based model reduces
to the SSTM.

In comparison with BDTM and DDTM, SSTM is featured by its natural capability in characterizing the time-varying
property of $\Theta$. This feature renders SSTM to be a better choice for characterizing dynamically changing sensing
environments in many WSN applications such as robust
sensor data fusion \cite{liu2015toward} and fault-tolerant event
detection \cite{wang2017online}.
\section{Subjective logic based trust model (SLTM)}\label{sec:sltm}
Subjective logic operates on subjective beliefs about an event or subject of interest \cite{josang1997artificial}, wherein a belief is
represented by the notion of opinion. In contrast with classical probability calculus that uses standard first-order probability representation,
subjective logic employs second-order probabilities that is equivalent to a pdf over a first-order probability variable.

Assume that the trustor has an opinion on its trust in the trustee.
Such an opinion is translated into degrees of belief or disbelief
about the trustworthiness of the trustee in SLTM. The major feature
of SLTM is that it quantitatively measures degrees of ignorance in a
direct way. Borrowing the same notation from
\cite{josang1997artificial}, we use $b$, $d$ and $i$ respectively to denote
belief, disbelief, and ignorance. In SLTM, it is
assumed that
\begin{equation}\label{eqn:sl}
b+d+i=1, \{b,d,i\}\in[0,1]^3,
\end{equation}
and an opinion on trust can be uniquely described by the triplet $\pi=\{b,d,i\}$.
See Algorithm \ref{alg:SLTM} for a pseudo-code to implement trust inference based on SLTM.
\begin{algorithm}[tb]
\caption{A pseudo-code to implement trust inference based on SLTM}
\label{alg:SLTM}
\begin{algorithmic}[1] 
\STATE Set the initial subjective opinion about the trust via a ternary array $SO=\{b,d,i\}$ that satisfies Eqn.(\ref{eqn:sl}).
\FOR{$k$=1,2,\ldots}
\STATE Given a new observation $y_k$, which is in the form of a subjective opinion about the trust, construct another ternary array, denoted by $SO'=\{b',d',i'\}$, to model $y_k$;
\STATE Update the posterior subjective opinion about the trust by setting  $SO\leftarrow s(SO, SO')$, where $s$ denotes a subjective logic operator, the details about which are referred to literature on the subjective logic, e.g., \cite{josang1999algebra}.
\ENDFOR
\end{algorithmic}
\end{algorithm}

In \cite{ivanovska2017joint}, Ivanovska et al. present a method for computing the joint subjective opinion of multiple variables.
In \cite{pope2005analysis}, Pope and J{\o}sang develop a subjective logic based approach to the evaluation of competing hypotheses.
In \cite{ivanovska2016bayesian}, Ivanovska, J{\o}sang, and Sambo present an extension of Bayesian deduction to the framework of subjective logic. In \cite{josang2008conditional}, J{\o}sang focuses on the conditional reasoning in subjective logic,
whereby beliefs are represented as binomial or multinomial subjective opinions.
In \cite{josang2012interpretation} and \cite{josang2017multi}, the authors
address the issue of subjective logic based belief fusion, which consists of taking into account multiple sources of belief about a domain of interest. In particular,
the article \cite{josang2012interpretation} discusses the selection of belief fusion operators and suggests to consider the nature of the situation to be modeled in searching the most appropriate one. The paper \cite{josang2017multi} describes
cumulative and averaging multi-source belief fusion in the formalism of subjective logic, which represents generalizations of binary-source belief fusion operators.
In \cite{josang2016principles}, J{\o}sang and Kaplan introduce the term \emph{subjective networks}, which generalizes Bayesian networks by letting it be based on subjective logic instead of probability calculus.
In \cite{oren2007subjective}, Oren, Norman, and Preece introduce a subjective logic based argumentation framework, which is
primarily used for evidential reasoning.

In \cite{josang1999algebra}, J{\o}sang applies SLTM to determine the trustworthiness of agents that are responsible for key generation and distribution in open networks.
In \cite{alhadad2014trust}, Alhadad et al.
use subjective logic to express and deal with uncertainty in evaluating trust
in a graph composed of paths that have common nodes.
In \cite{josang2006trust,josang2008optimal}, J{\o}sang et al. present an SLTM based method for trust network analysis. In \cite{liu2011novel}, Liu et al. present a subjective logic based reputation model for blocking
selfish behaviors in MANETs.
%
\subsection{On relationships between SLTM and
GBT}\label{sec:relation_sltm} In this section, we discuss the
connections between SLTM and GBT. As aforementioned, the most important feature of SLTM is that it allows
to quantitatively measure the degree of ignorance in evaluating
whether a trustee is trusted or not. It models belief $b$, disbelief
$d$, and ignorance $i$ on the trustworthiness of the trustee by a
triplet $\pi=\{b,d,i\}$, which satisfies (\ref{eqn:sl}). In concept,
we can interpret $i$ as an intermediate state on the
`belief'-`disbelief' scale. Now let us suppose that, in the context
of DDTM, the interaction outcomes between the trustor (or advisor)
and the trustee can be represented by three degrees of belief on the
trustworthiness of the trustee, namely `belief' $b$, `ignorance'
$i$, and `disbelief' $d$. Then the trust parameter can be
represented as $\Theta\equiv[b,i,d]$.
Let the prior of $\Theta$ follow a Dirichlet distribution $D(\alpha_1,\alpha_2,\alpha_3)$. Then,
given an $n$-sequence $X$ of trials $(n=n_1+n_2+n_3)$ with 3 distinct outcomes, the posterior follows the Dirichlet distribution
\begin{equation}
D(\alpha_1+n_1,\alpha_2+n_2,\alpha_3+n_3),
\end{equation}
where $n_j$ counts the occurrences of the $j$th type outcome in the sequence $\textbf{X}$.
As is shown, a bijective mapping between subjective opinions leveraged in SLTM and evidence parameters of Dirichlet pdfs exists. This bijective mapping is confirmed in \cite{josang2012interpretation}. In \cite{ivanovska2016bayesian}, Ivanovska, J{\o}sang and Sambo also state that
\begin{quote}
 ``Every subjective opinion can be ``projected'' to a single probability distribution, called projected probability distribution which is an important characteristic of the opinion since it unifies all of its defining parameters.''
\end{quote}
This mapping has been exploited for translating from inference with subjective opinions to inference with a corresponding Dirichlet
pdf \cite{Kaplan2016Efficient,Kaplan2013Reasoning}.

To summarize, DDTM can also be applied to measure the degree of ignorance in evaluating the trustworthiness of a trustee, in a similar spirit as SLTM. Since DDTM can be regarded as a specific ad hoc implementation of the GBT framework, as discussed in Section
\ref{sec:relation_bdtm}, the connections between SLTM and the GBT framework is built by using the DDTM as a bridge between them.
\section{An overall comparison of all model types under survey}\label{sec:overview}
In this section, we attempt to give a big picture of relationships among the aforementioned model types. Table \ref{tab:comparison1} lists the data types to be processed, forms of the prior and the likelihood function adopted, by each model type under consideration. In this table, we show that adopting the GBT perspective allows us to process any type of observations, either discrete or continuously valued, and adopt any form of prior or likelihood functions, in a completely Bayesian manner. This prominent feature of the GBT perspective can extend the application scope of Bayesian-based trust models to wider areas. A further comparison for model types within the Bayesian paradigm is presented in Table \ref{tab:comparison2}. It shows that, compared with BDTM, DDTM, and SSTM, GBT perspective is more flexible and generic.  
\begin{table}
\caption{A comparison of the model types under survey in terms of the data type to be processed, forms of the prior and the likelihood function adopted}
\centering
\begin{tabular}{ccccc}
\toprule
Model Class & type of observation & Prior & Likelihood\\
\midrule
BDTM       & Binary         & Beta        & Binomial\\
DDTM       & multiple typed & Dirichlet   & multinomial\\
SSTM       & Continuous & truncated normal  & any form\\
SLTM       & Three-valued & a ternary array $\{b,d,i\}$ & N/A \\
GBT        & any type &  any form  & any form \\
\bottomrule
\end{tabular}
\label{tab:comparison1}
\end{table}
\begin{table}
\caption{Properties of Bayesian trust models}
\centering
\begin{tabular}{ccccc}
\toprule
item & BDTM & DDTM & SSTM & GBT \\
\midrule
Conjugate prior required? & yes & yes & no & no \\
Support analytic inference? & yes & yes & no & depends \\
Explicitly model the evolution of trust?  & no & no & yes & yes \\
application-specific? & yes & yes & yes & no\\
\bottomrule
\end{tabular}
\label{tab:comparison2}
\end{table}
\section{Discussions on the GBT perspective}\label{sec:discuss}
In this section, we discuss the capabilities, limitations of the GBT perspective in
trust modeling, and then point out several open research issues to answer in the way to advance it to become a pragmatic infrastructure for analyzing intrinsic relationships between variants of trust models and for developing novel ones for trust evaluation.
\subsection{On capabilities of the GBT perspective}
As presented in Sections \ref{sec:relation_bdtm},
\ref{sec:relation_sstm}, and
\ref{sec:relation_sltm}, the GBT perspective provides a basic
theoretical tie that connects BDTM, DDTM, SSTM, and
SLTM together. It is shown that all these models cast into a common framework as special
implementation cases of the GBT perspective. This is due to the natural power the Bayesian philosophy owns in uncertainty quantification and the inherent capabilities possessed by the GBT perspective, which will be discussed as
follows.
\subsubsection{Capability to characterize the non-symmetry property of the trust}
It is an agreement in the literature that trust is not symmetric
\cite{cahill2003using,Teacy2008Hierarchical}, which implies that two
peers involved in an interaction may not necessarily have the same
trust in each other, even if they are presented with the same
evidence. This is because the trustor and trustee may not
necessarily interpret the outcome of each interaction in the same
way \cite{Teacy2008Hierarchical}.

The GBT perspective provides a natural way to take into account of
such aforementioned non-symmetry in the interpretation of
interaction outcomes.
 Consider again an interaction involving two agents 1 and 2. Denote the data yielded from this interaction as $y$.
 Then we can use two different likelihood functions, represented here as $p_{1\rightarrow 2}(y|\theta_2)$ and $p_{2\rightarrow 1}(y|\theta_1)$,
 to formulate the two different interpretations of $y$, from perspectives of agent 1 and 2, respectively. Here $\theta_1$ and $\theta_2$ denote agent 2's trust in agent 1 and agent 1's trust in agent 2, respectively.
\subsubsection{Capability to take account of the dynamic property of the trust}\label{sec:capability_dynamic}
Some trust models, like TRAVOS, assume that the behavior of agents
does not change over time \cite{teacy2005coping}, while in many cases agents may change their
behavior over time actively or passively. In particular, some agents
may have time-based behavioral strategies. The GBT perspective
provides at least two simple mechanisms for taking account of the
dynamic property of the trust. The first one is to characterize the
time-evolution law of trust directly through the state transition
function (\ref{eqn:trans}). It has been proved successful in a
number of applications
\cite{liu2016state,wang2017online,liu2015toward}. The other
mechanism is to design a likelihood function that can reflect the
effect of time on the trust value. For instance, in the likelihood
function, we can assign each observation a weight, whose value
depends on when this observation was generated.
A similar strategy of downweighting out-of-date ratings is presented in \cite{su2010pbtrust}.
\subsubsection{Capability to model the transitivity of trust}
The property of transitivity of trust is often employed in
establishing trust relationships between a pair of peers that have
never interacted before. In such situations, an intermediary peer,
also known as an advisor or third-party peer, can help if we allow
for transitive trust. The question needed to be addressed is that if
the trustor trusts the advisor to some extent, and the advisor has
past experience in interacting with the trustee, then to what extent
the trustor should trust the trustee. This notion of transitive
trust is very natural in many domains, e.g., when buyers of a product
recommend this product to new buyers and audience of a movie
recommend this movie to new audience. We can handle this
transitivity issue flexibly and efficiently when adopting the
GBT perspective. Suppose that an advisor provide historical data about the trustee to the trustor.
From the trustor's point of view, the observation it receives consists of two parts, namely the trustor's
trust $\theta_a$ in the advisor, and the
historical data $X$ about the trustee. We can design an appropriate
likelihood function $p(X|\theta,\theta_a)$ to take account of the
transitivity issue, while the form of the likelihood function is dependant on
application domains. Suppose that the BDTM is under use, and $X$ is
an $n$-sequence of $m$ successes, then we have $p(X|\theta,\theta_a)=\theta^{\theta_am}(1-\theta)^{\theta_a(n-m)}$.
Assume that the prior pdf follows $B(\alpha,\beta)$, then the posterior pdf becomes $B(\alpha+\theta_am,\beta+\theta_a(n-m))$.
It is clear that the value $\theta_a$ determines to what extent the data $X$ influences
the posterior. If $\theta_a=0$, which means that the advisor is
totally distrusted, then the posterior is equivalent with the prior,
indicating that the data provided by this advisor brings no new
information on the trustworthiness of the trustee.
\subsubsection{A potential capability to deal with data sparsity}\label{sec:sparsity}
The issue of data sparsity often appears in a very large network. For example, for some large e-commerce sites,
the number of sellers and buyers is so large that a buyer can hardly meet the same seller \cite{Wang03bayesiannetwork}.
In \cite{Resnick2002}, Resnick and Zeckhauser conduct an empirical study using eBay transaction data within a time period of five-month. They show that 89.0\% out of all seller-buyer pairs conducted only one transaction and 98.9\% conducted less than four.
It implies that the seller's trustworthiness is mostly evaluated based on a single transaction record by the buyers.
In \cite{zhou2007powertrust}, Zhou and Hwang also confirm that it is extremely rare to find a node with a
large number of feedbacks in a dynamically growing P2P network, because most of the network nodes only receive a few
feedbacks.

Most approaches work under a basic assumption that two agents can repeatedly interact with each other, which may be the case for a small-size network.
Because of the aforementioned data sparsity issue, these approaches are likely to fail to provide a satisfactory performance for large-scale networks.

As discussed in previous sections, the GBT perspective provides a
theoretically sound way to analyze and integrate various typed
observations, such as binary data, multi-valued data, subjective opinions, based on a
consistent modeling language. To conquer problems resulted from data
sparsity, we may leverage the power of GBT to take into account of social data, e.g.,
roles of and relationships between participating peers, and subject
opinions of domain experts, in forming a more accurate trust assessment.
\subsection{On limitations of GBT in trust modeling}\label{sec:limitation}
As a data-driven modeling framework, the success of a GBT based solution may depend
crucially on the quality and richness of data and domain knowledge
available. For example, if some third-party advisors provide unfair
ratings of the trustee, then it requires that there be enough peers
who offer honest ratings to override the effect of unfair ones.
Provided that there is no other social metadata or domain knowledge
available for use, a GBT based solution is likely to fail to produce an accurate
trust evaluation due to shortage of honest rating
data.

As discussed in Section \ref{sec:sparsity}, GBT has the potential
capability to handle data sparsity, while its working requires
preexisting social relationships among the network peers or domain
knowledge about the network. But they may not exist in practical
cases. For such cases, concrete security mechanisms, such as methods
based on credentials and policies, may be the only way to establish
a trust relationship between peers.

Another limitation of GBT lies in a heavy computation and memory
burden caused by running PF in scenarios that demand lightweight algorithms. As a sequential sampling based
approximation method, the computation and memory burden of PF is
linearly dependant on the sample size. Thus, to reduce computation
burden, the sample size needs to be controlled elaborately,
otherwise, the accuracy of PF estimation will deteriorate severely.
So unless the PF theory itself makes progress, the application scope of GBT based solutions
are restricted. Another choice to get rid of heavy
computation burden of PF is to employ a conjugate pair of prior pdf and likelihood function, while this may lead to performance deterioration due to model mismatch. To
conquer the above limitation, the fundamental solution should come from
advances in semiconductor technology that can enable a network
agent to perform more computations and possess larger capacities in
the memory.
\subsection{Open research questions}
\subsubsection{How to set an appropriate initial prior pdf of trust to start the computation engine}
Running of any algorithm in a Bayesian flavor requires a pre-setting
of the initial pdf, namely, $p(\theta_0)$ in the GBT framework. On
one side, we hope that this prior pdf can reflect accurately the
trustor's prior knowledge on the trustworthiness of the trustee,
especially for cases in which the trustor has inadequate observations in generating the posterior estimate from the prior. On the other side, we hope that a prior pdf conjugate with the
likelihood function can be used since this can make the posterior inference analytically tractable.
Thus, how to balance the requirement of representation accuracy and
that of computational tractability in setting the prior pdfs remains the first open question to be resolved.
\subsubsection{How to characterize multi-faceted and differentiated sides of trust in a model}\label{sec:multifaceted}
As a complicated concept, trust itself usually has multi-faceted and differentiated sides
in practice. For instance, an overall trust of a WSN system can be decomposed into two
components, termed communication trust and data trust, representing two different
aspects of the system trust \cite{momani2008bnwsn}.
The question is how to evaluate the trustworthiness of a system by taking account of the trustworthiness of the subsystems or atomic
components and the uncertainty associated with this information \cite{Ries2011}. The consideration of the trustworthiness of the subsystems
may be independent from how these trust values are assessed.
The challenge lies in that, by piecing together trustworthiness of the separate entities that compose a system, we usually
can not get a result that describes the trustworthiness of the whole system.
Indeed, one of the major factors that influence the trust in a system is the architecture of the system,
which can not be reflected by trustworthiness of the separate entities.
In spite of a few application-specific attempts in the literature, e.g.,
methods based on the Bayesian network \cite{Wang03bayesiannetwork,wang2003bayesian,Wang03trustand},
there lacks a generic strategy or theory to address the above
question, especially for complex network systems.
\subsubsection{How to connect trust with decision and action}
The trust models provide a formal language for discussing interactions between agents, while, given a trust model, a natural question remains unaddressed, namely, in general terms, how to choose decisions or actions for the trustor based on the result of
trust inference.
In a principled way, we can resort to decision theory \cite{berger2013statistical},
which states that a rational agent should always act to maximize its expected utility (EU).
We use $U: \Theta\rightarrow\mathbb{R}$ to denote a utility function, for which higher values indicate more preferred outcomes, and vice versa, then EU is calculated as follows
\begin{equation}
\mbox{EU}=\int U(\Theta)p(\Theta)d\Theta.
\end{equation}
An appropriate utility function is required to be prescribed beforehand according to the trustor's domain-specific goals and
preferences.
\subsubsection{How to characterize trust transfer across different domains in a model}
The phenomenon of trust transfer across domains is common in real life.
For example, a middle school student who is proficient in math is likely to be also
proficient in physics. It indicates that our trust in this student in terms of physics
can be transferred from that in terms of math. While the notion of trust transfer is intuitively easy to comprehend,
it has not been modeled formally in the literature.
We argue that trust transfer is a concept that is closely related with
the notion of multi-faceted and differentiated trust since trust in terms of a specific domain can be
interpreted as one aspect of the system trust, which is defined across all domains considered.
\section{Conclusion}\label{sec:conclude}
This paper surveys trust models from a Bayesian perspective by
systematizing the knowledge from a long list of trust-related
research articles. Specifically, a general perspective for trust modeling and evaluation termed GBT is highlighted. It is
shown that all models under survey can cast into a GBT based framework. The connections from each surveyed model to the GBT framework are clarified. Both capabilities and
limitations of the GBT perspective are discussed, and a number of open
research questions are pointed out. Despite the complex character of trust, here we make an attempt to promote the accumulation rather than the fragmentation of theory and research on trust, hoping that this may be helpful for stimulating further and deeper research on trust modeling for networked agents.
\section*{Acknowledgement}
This work was partly supported by the National Natural Science Foundation of China (Nos.61571238 and 61572263).





\bibliographystyle{IEEEbib}
\bibliography{mybibliography}
\end{document}